\documentclass[reprint,twocolumn]{revtex4}
\usepackage{amsfonts}
\usepackage{amsmath}
\usepackage{amssymb}
\usepackage{charter}
\usepackage{graphicx}
\usepackage{subfigure}
\usepackage{graphicx}
\usepackage{epstopdf}
\usepackage{color}
\usepackage{tocvsec2}
\usepackage{enumerate}
\usepackage{graphicx}
\usepackage{dcolumn}
\usepackage{bm}

\begin{document}

\title{\textbf{\ Enhancing discrete-modulated continuous-variable
measurement-device-independent quantum key distribution via quantum catalysis%
}}
\author{ Wei Ye$^{1}$, Ying Guo$^{1}$, Yun Mao$^{2*}$, Hai Zhong$^{1\dag}$
and Liyun Hu$^{3\ddag}$}
\affiliation{$^{1}${\small School of Computer Science and Engineering, Central South
University, Changsha 410083, China}\\
$^{2}${\small School of Automation, Central South University, Changsha
410083, China}\\
$^{3}${\small Center for Quantum Science and Technology, Jiangxi Normal
University, Nanchang 330022, China}\\
$^{*\dag \ddag }$Corresponding authors:
maoyun3106@sina.com,zhonghai@csu.edu.cn and hlyun2008@126.com}

\begin{abstract}
The discrete modulation can make up for the shortage of transmission
distance in measurement-device-independent continuous-variable quantum key
distribution (MDI-CVQKD) that has an unique advantage against all
side-channel attacks but also challenging for the further performance
improvement. Here we suggest a quantum catalysis (QC) approach for enhancing
the performance of the discrete-modulated (DM) MDI-CVQKD in terms of the
achievable secret key rate and lengthening the maximal transmission
distance. The numerical simulation results show that the QC-based MDI-CVQKD
with discrete modulation that involves a zero-photon catalysis (ZPC)
operation can not only obtain a higher secret key rate than the original DM
protocol, but also contributes to the reasonable increase of the
corresponding optimal variance. As for the extreme asymmetric and symmetric
cases, the secret key rate and maximal transmission distance of the
ZPC-involved DM MDI-CVQKD system can be further improved under the same
parameters. This approach enables the system to tolerate lower
reconciliation efficiency, which will promote the practical implementations
with state-of-art technology.
\end{abstract}

\maketitle

\section{Introduction}

Quantum key distribution (QKD) \cite{1,2,3,4}, as one of the most mature
applications in cryptography, aims to share secret keys between two distant
legitimate users (Alice and Bob) even in the presence of an eavesdropper
(Eve) \cite{5,6,7}. Thanks to the usage of efficient detection schemes,
including the homodyne and heterodyne detections, the continuous variable
(CV) QKD systems \cite{8,9,10,11,57} not only promise high key rates, but
also can be easily compatible with the current optical communication
technologies. In particular, since the Gaussian-modulated (GM) CVQKD
protocol \cite{9} has been theoretically proved to be secure against
collective attacks \cite{12} and coherent attacks \cite{13}, this protocol
paves the way for the development of commercial applications. For example,
the field tests of the CVQKD system over 50 km commercial fiber have been
achieved in recent years \cite{14}. Whereas, in practical implementations,
the existence of imperfect detectors may lead to the potential security
loopholes that could be successfully exploited by Eve to take some attack
strategies, e.g., the local oscillator calibration attack \cite{15}, the
wavelength attack \cite{16}, and the detector saturation attack \cite{17},
which are still problems for realizing the practical CVQKD system.

In response to these problems, the measurement-device-independent (MDI)
CVQKD protocols were proposed \cite{20,21,22,23,24,25,26,27,56}, where the
secret keys between Alice and Bob can be extracted by relying on the
measurement results of an untrusted third party (Charlie). This protocol can
be immune to all attacks on detectors, and its practical security proof has
been analyzed in both the finite-size effect \cite{25,26} and the composable
security \cite{27}. However, in contrast to its discrete-variable (DV)
counterpart \cite{18,19}, this MDI-CVQKD has limitations in the transmission
distance, which is not sufficient for the requirements of the long-distance
communication. Therefore, it is an urgent task to lengthen the maximal
transmission distance with underlying technology.

Up till now, most of investigations have focused on performance improvement
of the MDI-CVQKD system by means of quantum operations \cite{33,34,35}, such
as the phase-sensitive optical amplifier and the photon subtraction. To be
more specific, using a phase-sensitive optical amplifier to compensate the
imperfection of the Bell-state measurement (BSM) implemented by Charlie can
improve the performance of the MDI-CVQKD system \cite{33}. In addition, the
photon subtraction operations \cite{34,35} have been used for lengthening
the maximal transmission distance of the MDI-CVQKD system, and meanwhile
these photon-subtracted operations can be emulated by the non-Gaussian
postselection in order to circumvent the complexity of configurations.
Despite existing the aforementioned advantages, the success probability of
implementing photon-subtracted operations is less than $0.25$, which would
lead to the limited performance improvements \cite{36,37}. To eliminate this
drawback, recently, the quantum catalysis \cite{36,37,38,39,40,41} has been
viewed as another useful method to extend the maximal transmission distance
of the MDI-CVQKD system \cite{42}, compared with the single-photon
subtraction case. Unfortunately, a major problem, common to the
aforementioned GM CVQKD, is that the reconciliation efficiency $\beta $ is
low, especially on the long-distance transmission with a low signal-to-noise
ratio.

To solve this problem, there are two approaches. One is to design effective
reconciliation code, such as low-density parity-check code \cite{43} and
multidimensional reconciliation code \cite{44}, but this kind of design has
a higher cost of hardwares and is hard to be realized in practice, and the
alternative is to employ the discrete-modulated (DM) schemes \cite%
{28,29,30,31,32,51,54,55}, including the four-state protocol and the
eight-state protocol, which allows the distribution of secret keys over long
distances in the regime of low signal-to-noise ratio with an efficient
reconciliation efficiency. Recently, it has been shown \cite{32} that the
DM-based MDI-CVQKD under the extreme asymmetric case rather than the
symmetric case has an advantage over the GM-based one with respect to the
maximal transmission distance. Motivated by the advantages of the DM schemes
with an efficient reconciliation efficiency, in this paper, we put forward a
feasible method using a zero-photon catalysis (ZPC) operation in order to
further enhance the performance of MDI-CVQKD with discrete modulation. Our
results under the same accessible parameters show that, as for the extreme
asymmetric and symmetric cases, the usage of the ZPC operation on MDI-CVQKD
with discrete modulation can not only give birth to the high secret key
rates, but also contributes to the reasonable increase of the corresponding
optimal variance. Furthermore, the maximal transmission distance can be
lengthened when comparing with the traditional DM scheme. More
interestingly, our protocol enables the MDI-CVQKD with discrete modulation
to tolerate lower reconciliation efficiency.

The rest of this paper is structured as follows. In Sec. II, we first review
the entanglement-based (EB) version of the original DM MDI-CVQKD protocol
and then demonstrate the effects of the ZPC operation on MDI-CVQKD with
discrete modulation in detail. In Sec. III, according to the optimality of
Gaussian attack, we derive the asymptotic secret key rate of the
ZPC-involved MDI-CVQKD with discrete modulation. In Sec. IV, the simulation
results and discussions are provided. Finally, our conclusions are drawn in
Sec. V.

\begin{figure*}[tph]
\label{Fig1} \centering \includegraphics[width=1.2\columnwidth]{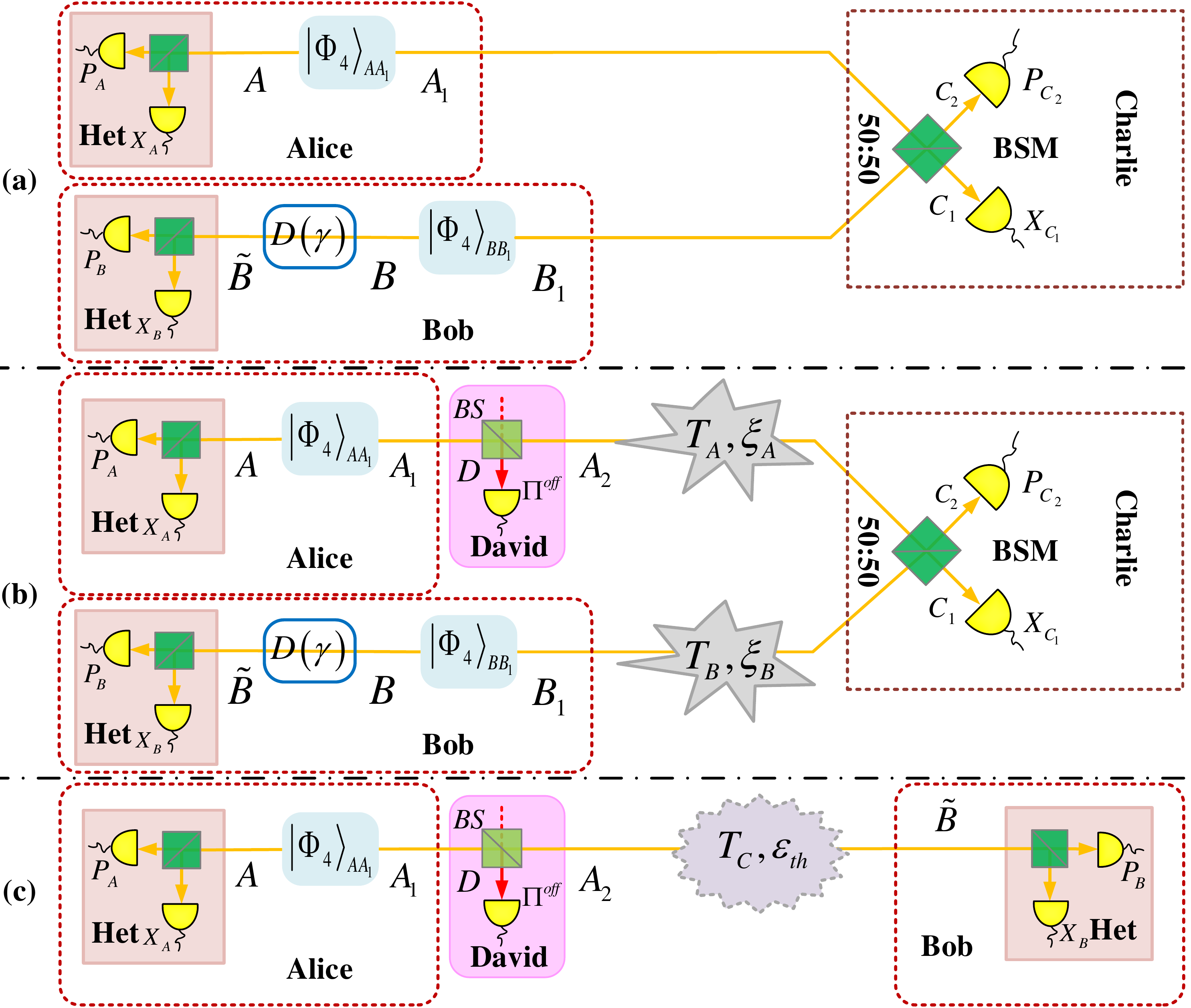}
\caption{{}(Color online) (a) EB version of the original MDI-CVQKD with
discrete modulation. (b) EB version of the ZPC-involved MDI-CVQKD with
discrete modulation. (c) Equivalent one-way CVQKD protocol with the ZPC
operation under the assumption that Bob's prepared state $\left \vert \Phi
_{4}\right \rangle _{BB_{1}}$ and displacement except for heterodyne
detection are untrusted. Both $\left \vert \Phi _{4}\right \rangle _{AA_{1}}$
and $\left \vert \Phi _{4}\right \rangle _{BB_{1}} $ are the two-mode
entangled states. 50:50 is a balanced beam splitter. BS is a beam splitter
with a transmittance $T$. BSM is a Bell-state measurement. Het and Hom are
respectively a heterodyne detection and a homodyne detection. $\left \{
X_{j},P_{j}\right \} \left( j=A,B\right) $ is the outcome results of the
heterodyne detection. $X_{C_{1}}$ (or $P_{C_{2}}$) is the outcome results of
the BSM. $D\left( \protect \gamma \right) $ is a displacement operator. $\Pi
^{off}=\left \vert 0\right
\rangle _{D}\left \langle 0\right \vert $ is a
projective operator. $T_{j}$ and $\protect \xi _{j}$ ($j=A,B$) are
respectively the quantum channel transmittance and the excess noise. $T_{C}$
and $\protect \varepsilon _{th}$ are respectively the equivalent quantum
channel transmittance and excess noise of the equivalent one-way protocol.}
\end{figure*}

\section{ZPC-involved MDI-CVQKD protocol with discrete modulation}

The EB version of MDI-CVQKD with discrete modulation is shown in Fig. 1(a),
where Alice and Bob respectively generate a two-mode entangled state $%
\left
\vert \Phi _{4}\right \rangle _{AA_{1}}$ and $\left \vert \Phi
_{4}\right
\rangle _{BB_{1}}$, and then both of them respectively hold
modes $A$ and $B$ while sending modes $A_{1}$ and $B_{1}$, along with the
quantum channels of lengths $L_{AC}$ and $L_{BC}$, to an untrusted third
party Charlie who proceeds to interfere with the incoming modes $A_{1}$ and $%
B_{1}$ at a symmetric beam splitter, and attains two output modes $C_{1}$
and $C_{2} $. After the $x$ ($p$) quadrature of mode $C_{1}$ ($C_{2}$) is
measured by the BSM, Charlie proclaims the measured results $\left \{
X_{C_{1}},P_{C_{2}}\right \} $. Subsequently, Bob modifies mode $B$ to $%
\widetilde{B}$ by using a displacement operation $D\left( \gamma \right) $
according to the public results $\left \{ X_{C_{1}},P_{C_{2}}\right \} ,$
where $\gamma =g\left( X_{C_{1}}+iP_{C_{2}}\right) $ with the gain $g$ of a
displacement operation. Through heterodyne detections, Alice and Bob
respectively measure mode $A$ and $\widetilde{B}$ to attain the outcome
results $\left \{ X_{A},P_{A}\right \} $ and $\left \{ X_{B},P_{B}\right \}
, $ and finally both of them have to implement the postprocessing to get a
string of secret keys.

In what follows, we focus on the ZPC-involved MDI-CVQKD with discrete
modulation. For the sake of discussions, here we take the EB version of the
four-state protocol into account. As shown in Fig. 1(b), in the Alice's
station, Alice prepares for the two-mode entangled state $\left \vert \Phi
_{4}\right \rangle _{AA_{1}}$ on modes $A$ and $A_{1} $ with a modulated
variance $V_{M}=2\alpha ^{2}=V_{A}-1$ given by%
\begin{align}
\left \vert \Phi _{4}\right \rangle _{AA_{1}}& =\underset{l=0}{\overset{3}{%
\sum }}\sqrt{\lambda _{l}}\left \vert \phi _{l}\right \rangle _{A}\left
\vert \phi _{l}\right \rangle _{A_{1}}  \notag \\
& =\frac{1}{2}\underset{l=0}{\overset{3}{\sum }}\sqrt{\lambda _{l}}\left
\vert \varphi _{l}\right \rangle _{A}\left \vert \alpha _{l}\right \rangle
_{A_{1}},  \label{1}
\end{align}%
where the non-Gaussian orthogonal state $\left \vert \varphi
_{l}\right
\rangle _{A}$ on mode $A$ is expressed as
\begin{equation}
\left \vert \varphi _{l}\right \rangle _{A}=\frac{1}{2}\underset{j=0}{%
\overset{3}{\sum }}e^{i\left( 2l+1\right) j\pi /4}\left \vert \phi
_{j}\right \rangle _{A},\left( l=0,1,2,3\right) ,  \label{2}
\end{equation}%
with
\begin{equation}
\left \vert \phi _{l}\right \rangle =\frac{e^{-\alpha ^{2}/2}}{\sqrt{\lambda
_{l}}}\underset{m=0}{\overset{3}{\sum }}\frac{\left( -1\right) ^{m}\alpha
^{4m+l}}{\sqrt{\left( 4m+l\right) !}}\left \vert 4m+l\right \rangle ,
\label{3}
\end{equation}%
\begin{align}
\lambda _{0,2}& =\frac{e^{-\alpha ^{2}}}{2}\left[ \cosh \alpha ^{2}\pm \cos
\alpha ^{2}\right] ,  \notag \\
\lambda _{1,3}& =\frac{e^{-\alpha ^{2}}}{2}\left[ \sinh \alpha ^{2}\pm \sin
\alpha ^{2}\right] ,  \label{4}
\end{align}%
\ and $\left \vert \alpha _{l}\right \rangle _{A_{1}}=$ $\left \vert \alpha
e^{i(2l+1)\pi /4}\right \rangle $ ($l=0,1,2,3$) on mode $A_{1}$ is a
modulated coherent state with a positive number $\alpha $. Note that, one of
the states $\left \vert \alpha _{l}\right \rangle _{A_{1}}$with the same
probability $1/4$ is randomly sent to Charlie through an unsecure quantum
channel. Thus, the corresponding covariance matrix $\Gamma _{AA_{1}}$ of the
resending state $\left \vert \Phi _{4}\right \rangle _{AA_{1}}$ is%
\begin{equation}
\Gamma _{AA_{1}}=\left(
\begin{array}{cc}
XI_{2} & Z_{4}\sigma _{z} \\
Z_{4}\sigma _{z} & XI_{2}%
\end{array}%
\right) ,  \label{5}
\end{equation}%
where $I_{2}=$diag$\left( 1,1\right) ,\sigma _{z}=$diag$\left( 1,-1\right) ,$
and
\begin{equation}
X=1+2\alpha ^{2},~Z_{4}=2\alpha ^{2}\underset{k=0}{\overset{3}{\sum }}%
\lambda _{k-1}^{3/2}\lambda _{k}^{-1/2}.  \label{6}
\end{equation}

In order to lower requirements for device perfection of the quantum
catalysis, we assume that, for our protocol, an untrusted third party,
David, who is near Alice, controls the ZPC operation (magenta box) where a
vacuum state $\left \vert 0\right \rangle _{D}$ in an auxiliary mode $D$ is
injected into the beam splitter (BS) with a transmittance $T$, and
simultaneously an ideal on/off detector placed at the output port of mode $D$
is used for registering the same state $\left \vert 0\right \rangle _{D}$.
As shown in Refs. \cite{36,37}, the vacuum state $\left \vert
0\right
\rangle _{D} $ between the input and output ports seems to be
unchanged, but promotes the transformation of quantum states between modes $%
A_{1}$ and $A_{2}$. Moreover, this ZPC operation can be described as an
equivalent operator
\begin{equation}
\widehat{O}_{0}=\text{Tr}\left[ \Pi ^{off}B\left( T\right) \right] =\sqrt{T}%
^{b^{\dag }b},  \label{7}
\end{equation}%
where $B\left( T\right) =\exp \left[ \left( ad^{\dag }-a^{\dag }d\right)
\arccos \sqrt{T}\right] $ is a BS operator and $\Pi ^{off}=\left \vert
0\right \rangle _{D}\left \langle 0\right \vert $ is a projective operator
on mode $D$. In addition, when such an operation is applied to an arbitrary
coherent state $\left \vert \alpha \right \rangle $, one can obtain%
\begin{align}
\left \vert \psi \right \rangle _{out}& =\frac{\widehat{O}_{0}}{\sqrt{P_{d}}}%
\left \vert \alpha \right \rangle  \notag \\
& =\frac{\exp [-\frac{\left \vert \alpha \right \vert ^{2}}{2}\left(
1-T\right) ]}{\sqrt{P_{d}}}\left \vert \sqrt{T}\alpha \right \rangle ,
\label{8}
\end{align}%
where $P_{d}=\exp [-\left \vert \alpha \right \vert ^{2}\left( 1-T\right) ]$
presents the success probability of implementing such an operation.
Evidently, with the assistant of the ZPC, the conversion of quantum states
from $\left \vert \alpha \right \rangle $ to $\left \vert \sqrt{T}\alpha
\right \rangle $ can be achieved with the probability $P_{d}$. Thus, when
one mode $A_{1}$ of the states $\left \vert \Phi _{4}\right \rangle
_{AA_{1}} $ is sent by Alice, the travelling state after succeeding in the
ZPC operation can be expressed as $\left \vert \Phi _{4}\right \rangle
_{AA_{2}}=\sum_{l=0}^{3}\sqrt{\lambda _{l}}/2\left \vert \varphi
_{l}\right
\rangle _{A}\left \vert \widetilde{\alpha }_{l}\right \rangle
_{A_{2}}$ with a covariance matrix \
\begin{equation}
\Gamma _{AA_{2}}=\left(
\begin{array}{cc}
\widetilde{X}I_{2} & \widetilde{Z}_{4}\sigma _{z} \\
\widetilde{Z}_{4}\sigma _{z} & \widetilde{X}I_{2}%
\end{array}%
\right) ,  \label{9}
\end{equation}%
where $\widetilde{X}$ and $\widetilde{Z}_{4}$ can be obtained by replacing $%
\alpha $ with $\widetilde{\alpha }$ in Eq. (\ref{5}), and $\left \vert
\widetilde{\alpha }_{l}\right \rangle _{A_{2}}=\left \vert \widetilde{\alpha
}e^{i(2l+1)\pi /4}\right \rangle _{A_{2}}$ with $\widetilde{\alpha }=\sqrt{T}%
\alpha $. It should be noted that after performing the ZPC operation, the
actual modulated variance of the travelling state $\left \vert \Phi
_{4}\right \rangle _{AA_{2}}$ turns out to be $\widetilde{V}_{M}=T\left(
V_{A}-1\right) .$ At the Bob's station, for simplicity, we assume that the
variance of the two-mode entangled state $\left \vert \Phi
_{4}\right
\rangle _{BB_{1}}$ is the same as that of the state $\left \vert
\Phi _{4}\right \rangle _{AA_{1}}$, i.e., $V_{A}=V_{B}=V$ throughout this
paper. Besides, due to the fact that Alice and Bob carry out the same
discrete modulation, the way of attaining the covariance matrix $\Gamma
_{BB_{1}}$ is the same as the Eq. (\ref{5}).

\section{The secret key rate of the ZPC-involved MDI-CVQKD system}

\begin{figure*}[t]
\centering
\subfigure[]{
\centering
\includegraphics[width=0.47\linewidth]{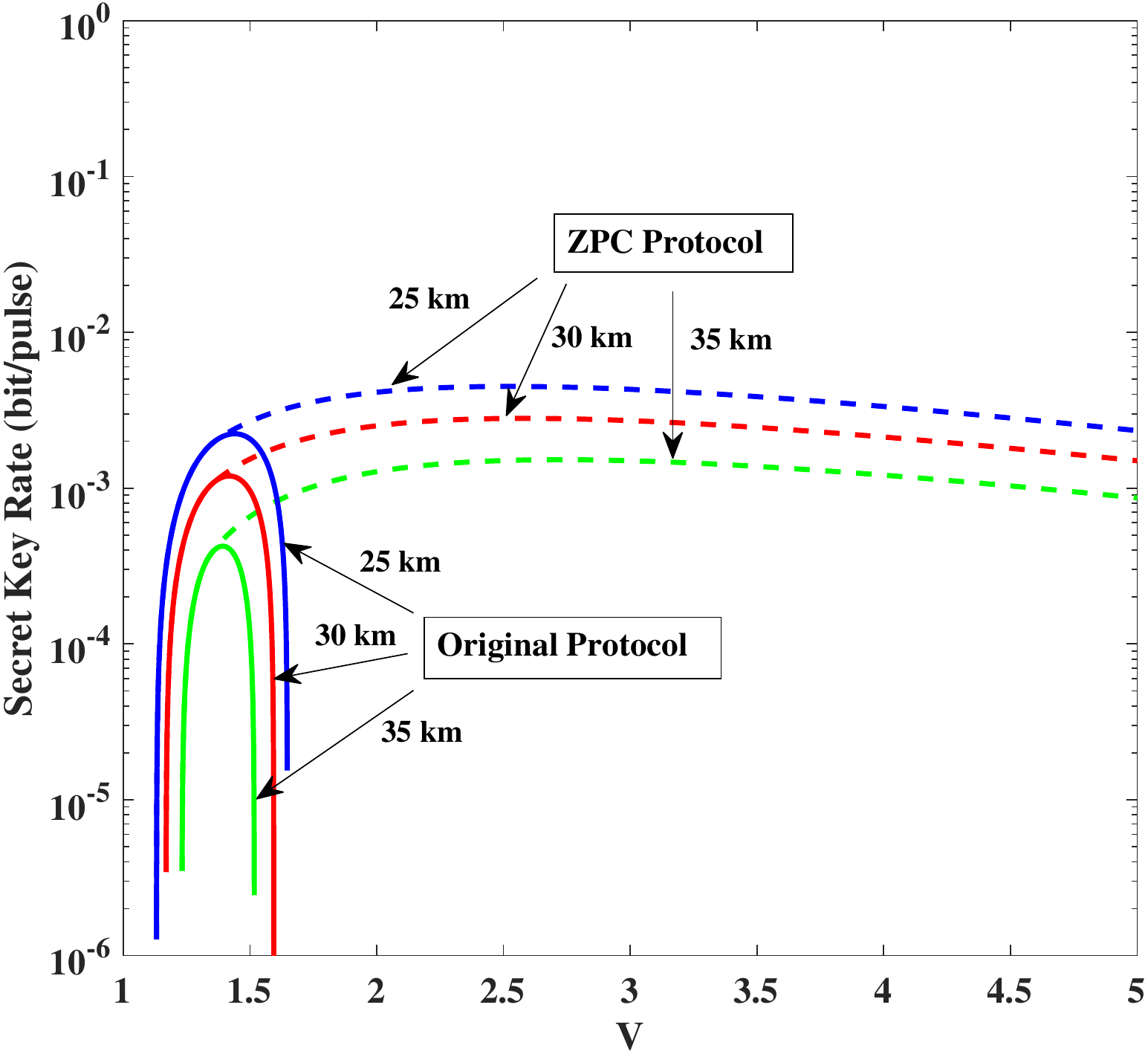}
\label{Fig2a}
}
\subfigure[]{
\includegraphics[width=0.46\linewidth]{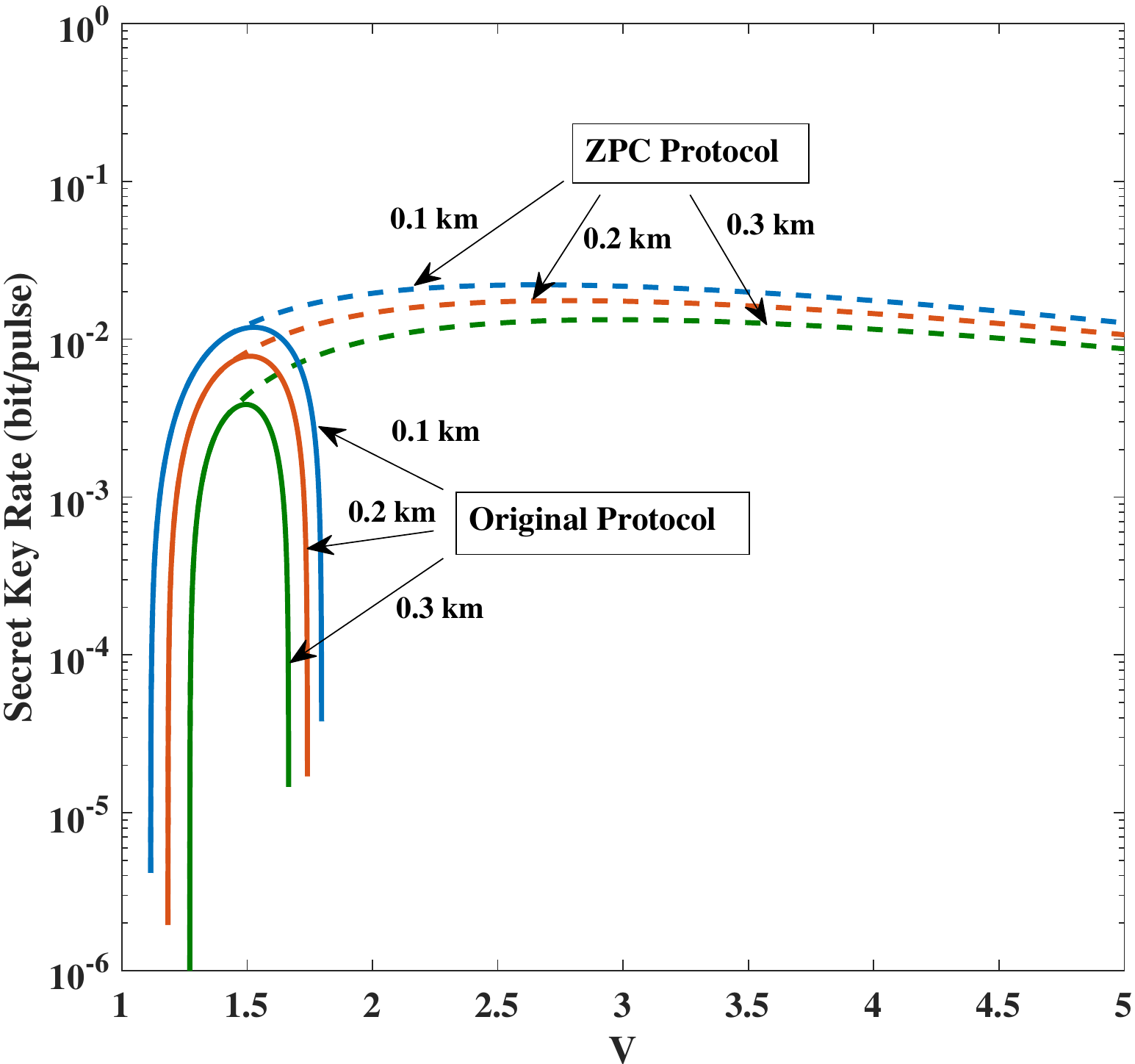}
\label{Fig2b}
}
\subfigure[]{
\includegraphics[width=0.47\linewidth]{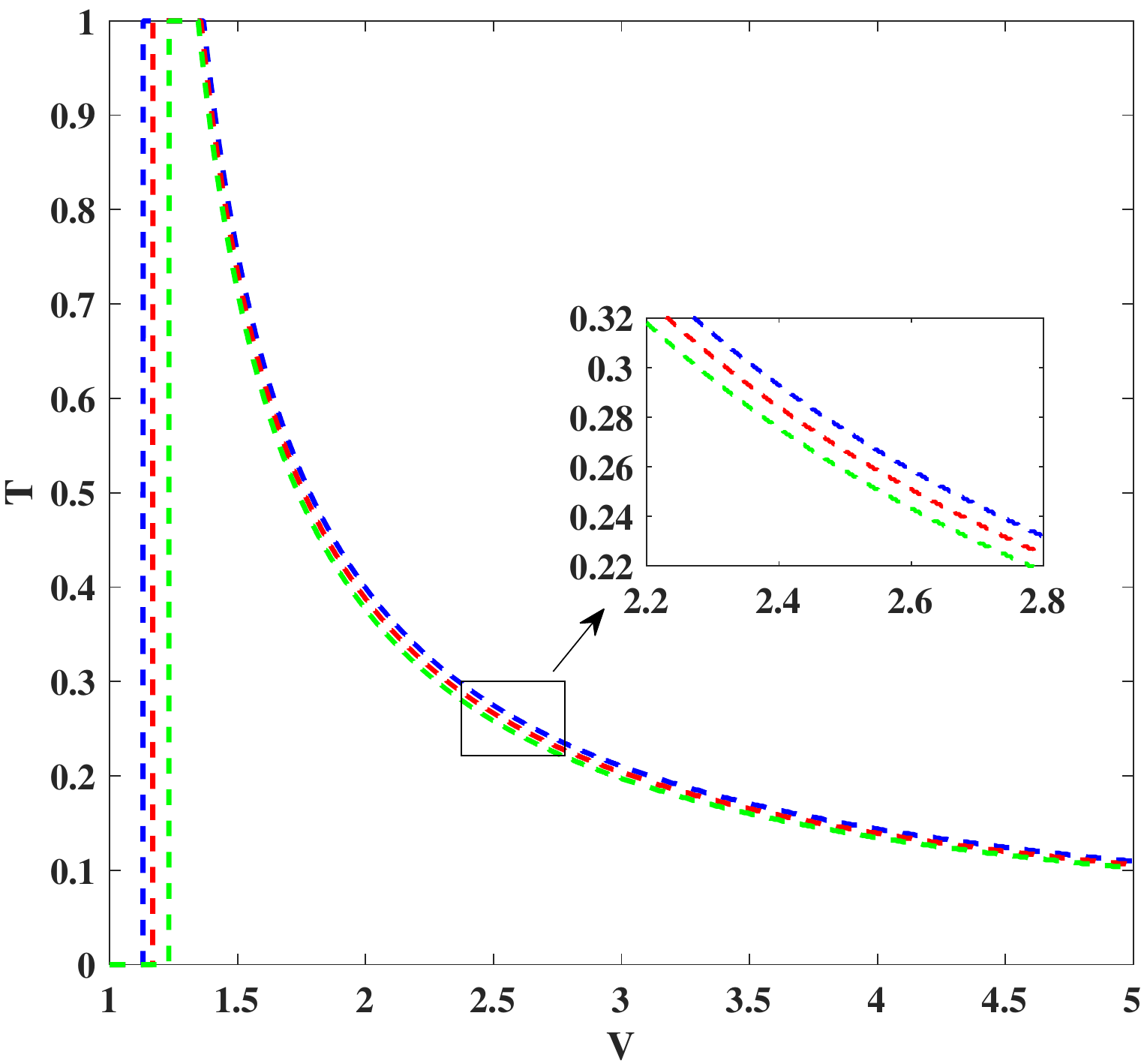}
\label{Fig2c}
}
\subfigure[]{
\includegraphics[width=0.46\linewidth]{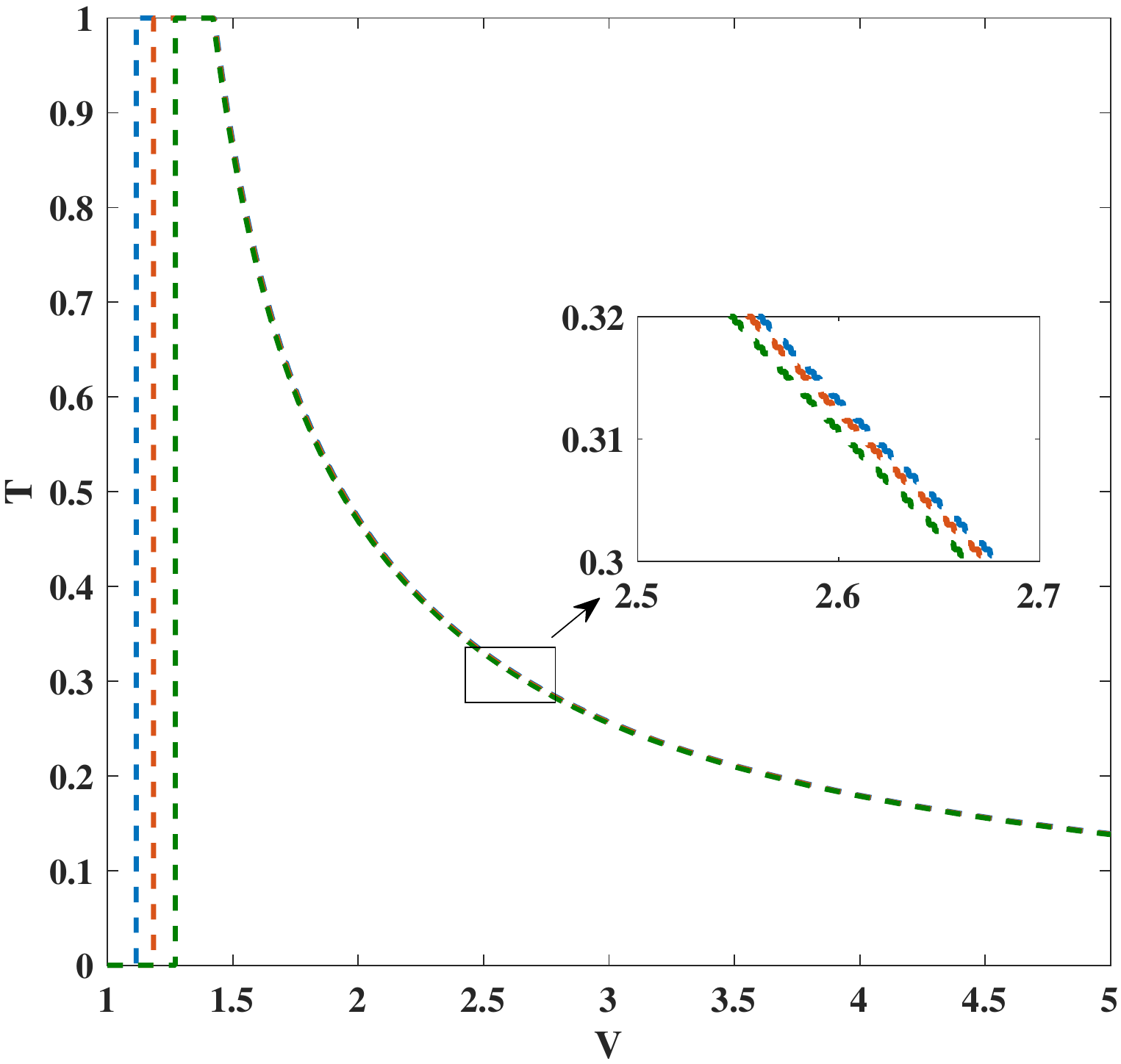}
\label{Fig2d}
}
\caption{{}(Color online) The secret key rate versus the variance $V$ for
(a) the extreme asymmetric case and (b) the symmetric case with transmission
distances $\{$$25$km, $30$km, $35$km$\}$ and $\{0.1$km $0.2$km, $0.3$ km$\}$
when optimizing over the transmittance $T$ shown in (c) and (d),
respectively. Note that ZPC Protocol and Original Protocol represent the
ZPC-involved MDI-CVQKD and the original MDI-CVQKD with discrete modulation.
Other parameters are given by reconciliation efficiency $\protect \beta =0.95$
and excess noise $\protect \xi _{j}=\protect \xi =0.002$ for $j\in \{A,B\}$}
\label{Fig2}
\end{figure*}

So far, we have established the schematic structure of the EB version of\
the ZPC-involved MDI-CVQKD system with discrete modulation. In this section,
for the reverse reconciliation algorithm, we pay attention to the derivation
of asymptotic secret key rates against one-mode collective Gaussian attacks
\cite{12,45} where two Markovian memoryless Gaussian quantum channels are
not related \cite{46}. Here we consider the results that the both channels
between Alice (or Bob) and Charlie are Gaussian under the condition of the
modulated variance $V_{M}=V-1<0.5$ \cite{28,29,30,31,32}, which arises from
the indistinguishability of $Z_{4}$ and $Z_{G}=\sqrt{V^{2}-1}$. It means
that the DM scheme can be viewed as the GM one when satisfying the condition
of $V_{M}<1.5$. In Fig. 1(b), we assume that the losses of both channels are
$\kappa =0.2$ dB/km, while $T_{A}=10^{-\kappa L_{AC}/10}$ ($%
T_{B}=10^{-\kappa L_{BC}/10}$) and $\xi _{A}$ ($\xi _{B}$) respectively
represent the transmittance and excess noise of the channel between Alice
(Bob) and Charlie. As referred to Refs. \cite%
{20,21,22,23,24,25,26,27,33,34,35}, the EB version of the MDI-CVQKD system
can be equivalent to a conventional one-way CVQKD system with the assumption
that both the state $\left \vert \Phi _{4}\right \rangle _{BB_{1}}$ and the
displacement operation $D\left( \gamma \right) $ in Bob's station are
untrusted except for the heterodyne detection, as shown in Fig. 1(c). For
the equivalent system, $T_{C}$ and $\varepsilon _{th}$ are respectively the
equivalent quantum channel transmittance and excess noise through the
relations%
\begin{align}
T_{C}& =\frac{g^{2}T_{A}}{2},  \notag \\
\varepsilon _{th}& =1+\chi _{A}+\frac{T_{B}}{T_{A}}\left( \chi _{B}-1\right)
\notag \\
& +\frac{T_{B}}{T_{A}}\left( \sqrt{\frac{2\left( V_{B}-1\right) }{g^{2}T_{B}}%
}-\sqrt{V_{B}+1}\right) ^{2},  \label{10}
\end{align}%
where $\chi _{j}=\left( 1-T_{j}\right) /T_{j}+\xi _{j}$ $(j=A,B)$. Further,
by setting $g^{2}=2\left( V_{B}-1\right) /\left[ \left( V_{B}+1\right) T_{B}%
\right] $, one can derive the minimum $\varepsilon _{th}$, i.e.,
\begin{equation}
\varepsilon _{th}=\frac{T_{B}}{T_{A}}\left( \chi _{B}-1\right) +1+\chi _{A}.
\label{11}
\end{equation}

Although the imperfection of the BSM in Charlie is an inescapable event,
such an imperfection can be compensated by optical preamplifiers \cite%
{35,50,53}. For simplicity, we consider the assumption of the perfection
detection for performance analysis. Consequently, the total noise of the
channel input in the shot-noise units (SNU) can be expressed as $\chi _{t}=$
$\left( 1-T_{C}\right) /T_{C}+\varepsilon _{th}$.

According to the aforementioned analysis, after performing the BSM and the
displacement operation, the covariance matrix $\Gamma _{A\widetilde{B}}$ of
the state $\left \vert \Phi _{4}\right \rangle _{A\widetilde{B}}$ is
\begin{align}
\Gamma _{A\widetilde{B}}& =\left(
\begin{array}{cc}
aI_{2} & c\sigma _{z} \\
c\sigma _{z} & bI_{2}%
\end{array}%
\right)  \notag \\
& =\left(
\begin{array}{cc}
\widetilde{X}I_{2} & \sqrt{T_{C}}\widetilde{Z}_{4}\sigma _{z} \\
\sqrt{T_{C}}\widetilde{Z}_{4}\sigma _{z} & T_{C}\left( \widetilde{X}+\chi
_{t}\right) I_{2}%
\end{array}%
\right) .  \label{12}
\end{align}%
We note that the ZPC operation belongs to a kind of Gaussian operation,
which makes it possible to derive the secret key rates by using the results
of the extremality of Gaussian quantum states \cite{47}. Thus, we have the
secret key rate of the ZPC-involved MDI-CVQKD system with discrete
modulation under reverse reconciliation against one-mode collective Gaussian
attacks, i.e.,%
\begin{equation}
K=P_{d}\left \{ \beta I\left( A\text{:}B\right) -\chi \left( B\text{:}%
E\right) \right \} ,  \label{13}
\end{equation}%
where the success probability $P_{d}$ has been given in Eq. (\ref{8}), $%
\beta $ denotes a reverse-reconciliation efficiency, $I\left( A\text{:}%
B\right) $ denotes the Shannon mutual information between Alice and Bob,
which can be derived as%
\begin{equation}
I\left( A\text{:}B\right) =\log _{2}\frac{a+1}{a+1-c^{2}/\left( Y+1\right) },
\label{14}
\end{equation}%
and $\chi \left( B\text{:}E\right) $ denotes the Holevo bound between Bob
and Eve. To obtain this Holevo bound, we assume that Eve is aware of the
untrusted third party David, and can purify the whole system $\rho _{A%
\widetilde{B}ED}$, so that
\begin{align}
\chi \left( B\text{:}E\right) & =S\left( E\right) -S\left( E|B\right)  \notag
\\
& =S\left( A\widetilde{B}\right) -S\left( A|\widetilde{B}^{m_{B}}\right) ,
\notag \\
& =\underset{i=1}{\overset{2}{\sum }}G\left[ \frac{\lambda _{i}-1}{2}\right]
-G\left[ \frac{\lambda _{3}-1}{2}\right] ,  \label{15}
\end{align}%
where $G\left[ x\right] =\left( x+1\right) \log _{2}\left( x+1\right) -x\log
_{2}x,S\left( A\widetilde{B}\right) $ is a function of the symplectic
eigenvalues $\lambda _{1,2}$ of $\Gamma _{A\widetilde{B}}$ given by
\begin{equation}
\lambda _{1,2}^{2}=\frac{1}{2}\left[ \vartheta \pm \sqrt{\vartheta
^{2}-4\zeta ^{2}}\right] ,  \label{16}
\end{equation}%
with $\vartheta =a^{2}+b^{2}-2c^{2}$ and $\zeta =ab-c^{2}$, and Eve's
condition entropy $S(A|\widetilde{B}^{m_{B}})$ based on Bob's measurement
result $m_{B}$, is a function of the symplectic eigenvalues $\lambda _{3}$
of $\Gamma _{A}^{m_{B}}=aI_{2}-c\sigma _{z}\left( YI_{2}+I_{2}\right)
^{-1}c\sigma _{z}$, which is given by
\begin{equation}
\lambda _{3}=\frac{a\left( b+1\right) -c^{2}}{b+1}.  \label{17}
\end{equation}

\begin{figure*}[t]
\centering
\subfigure[]{
\centering
\includegraphics[width=0.47\linewidth]{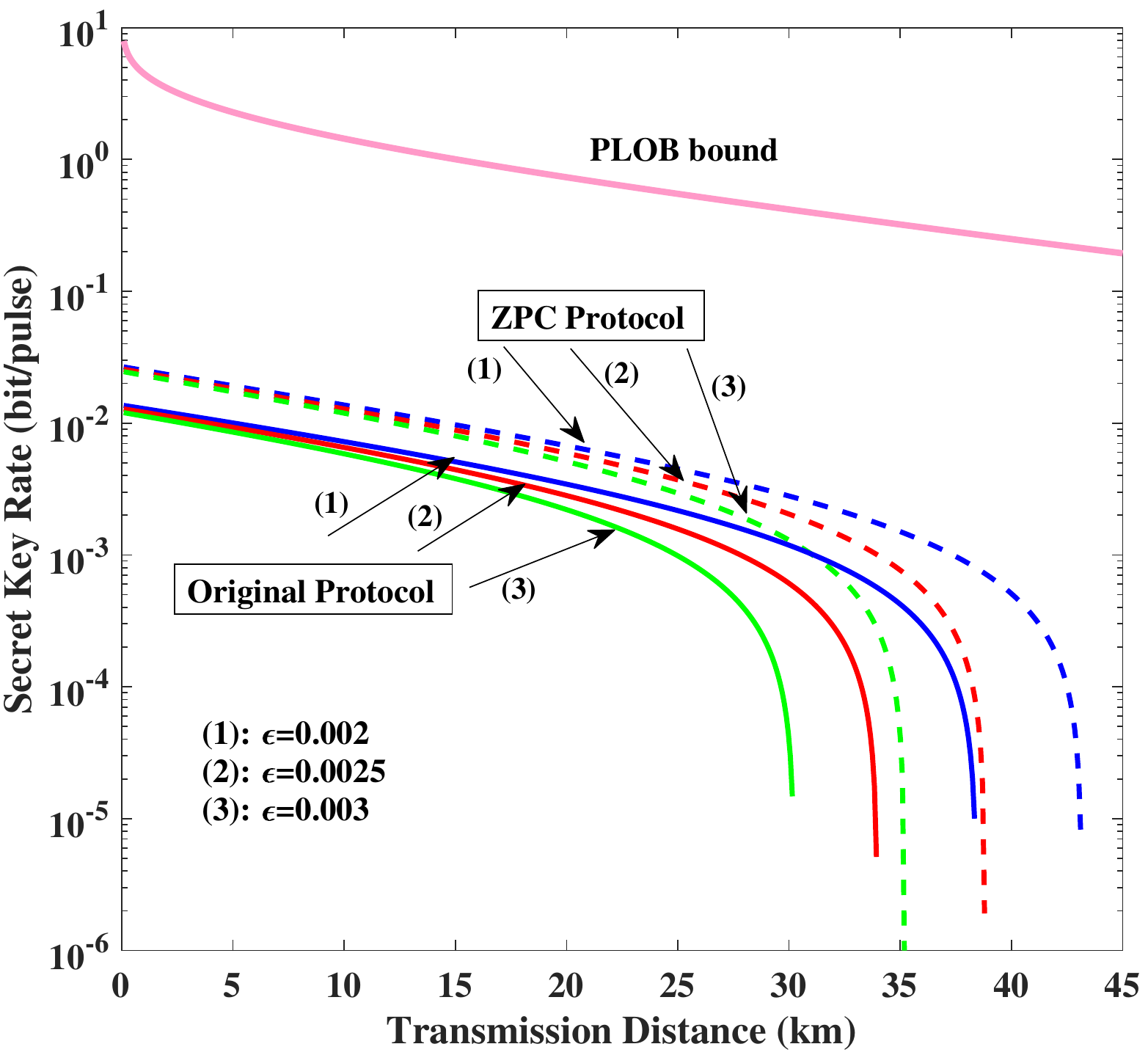}
\label{Fig3a}
}
\subfigure[]{
\includegraphics[width=0.46\linewidth]{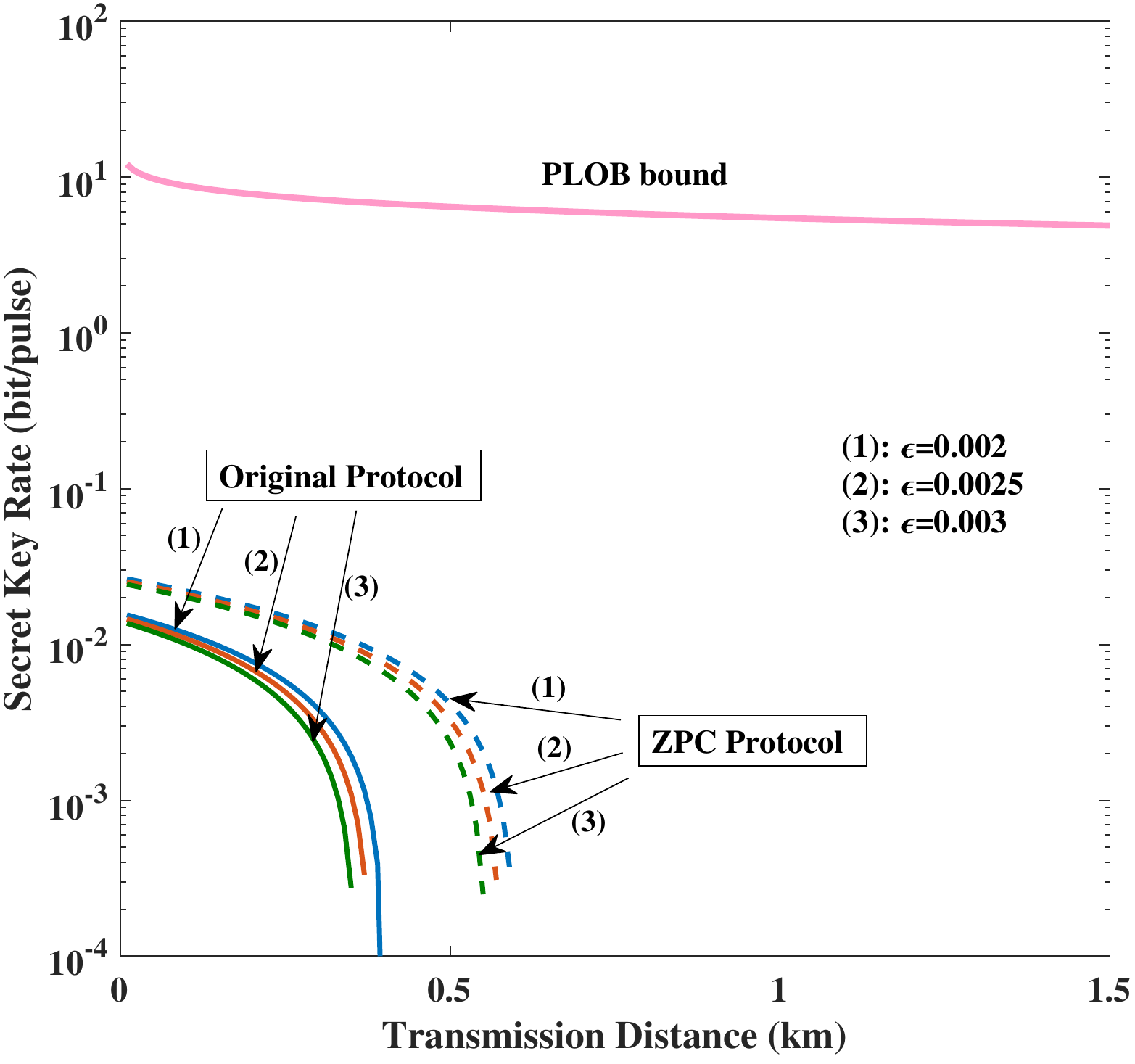}
\label{Fig3b}
}
\subfigure[]{
\includegraphics[width=0.47\linewidth]{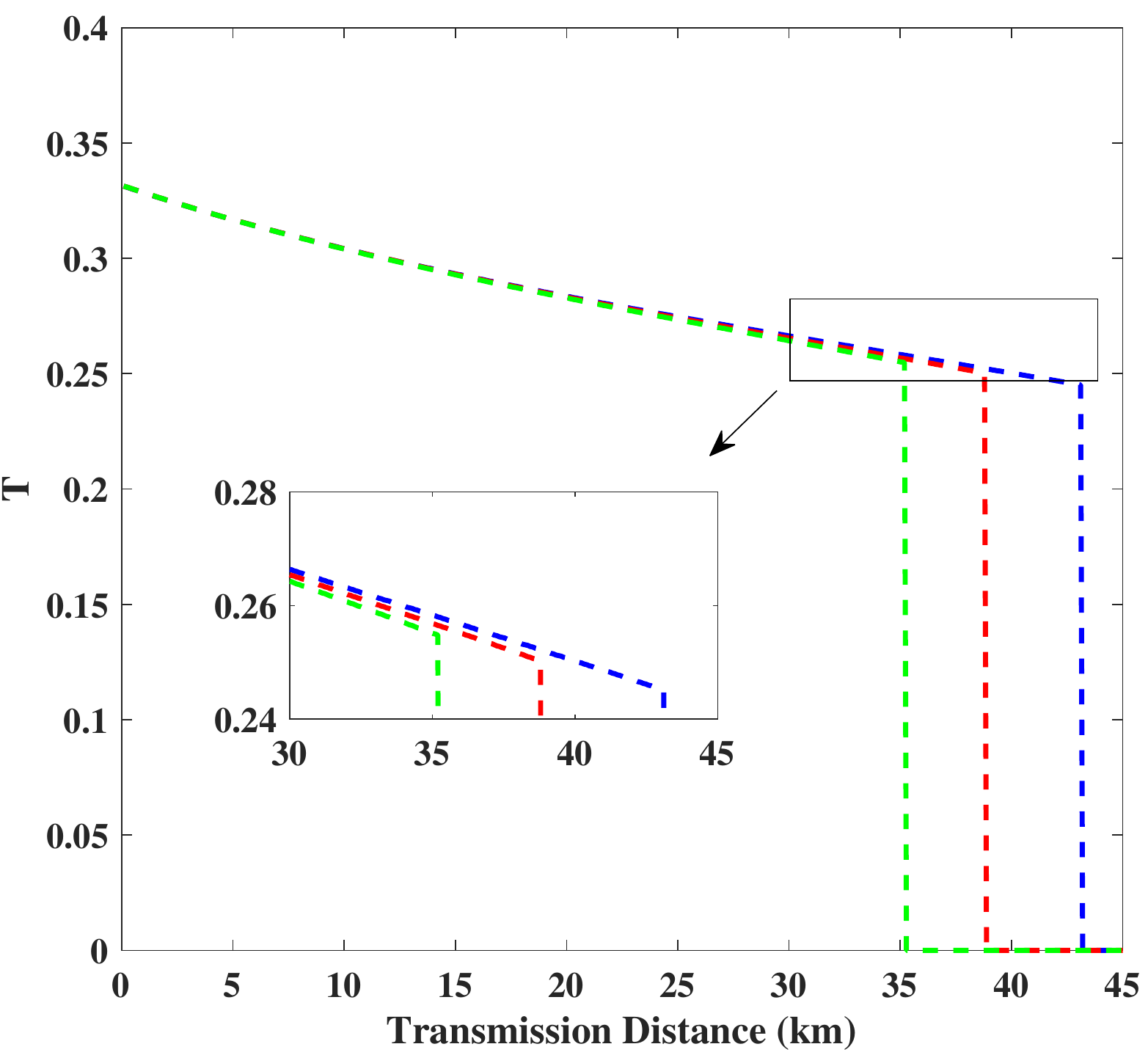}
\label{Fig3c}
}
\subfigure[]{
\includegraphics[width=0.46\linewidth]{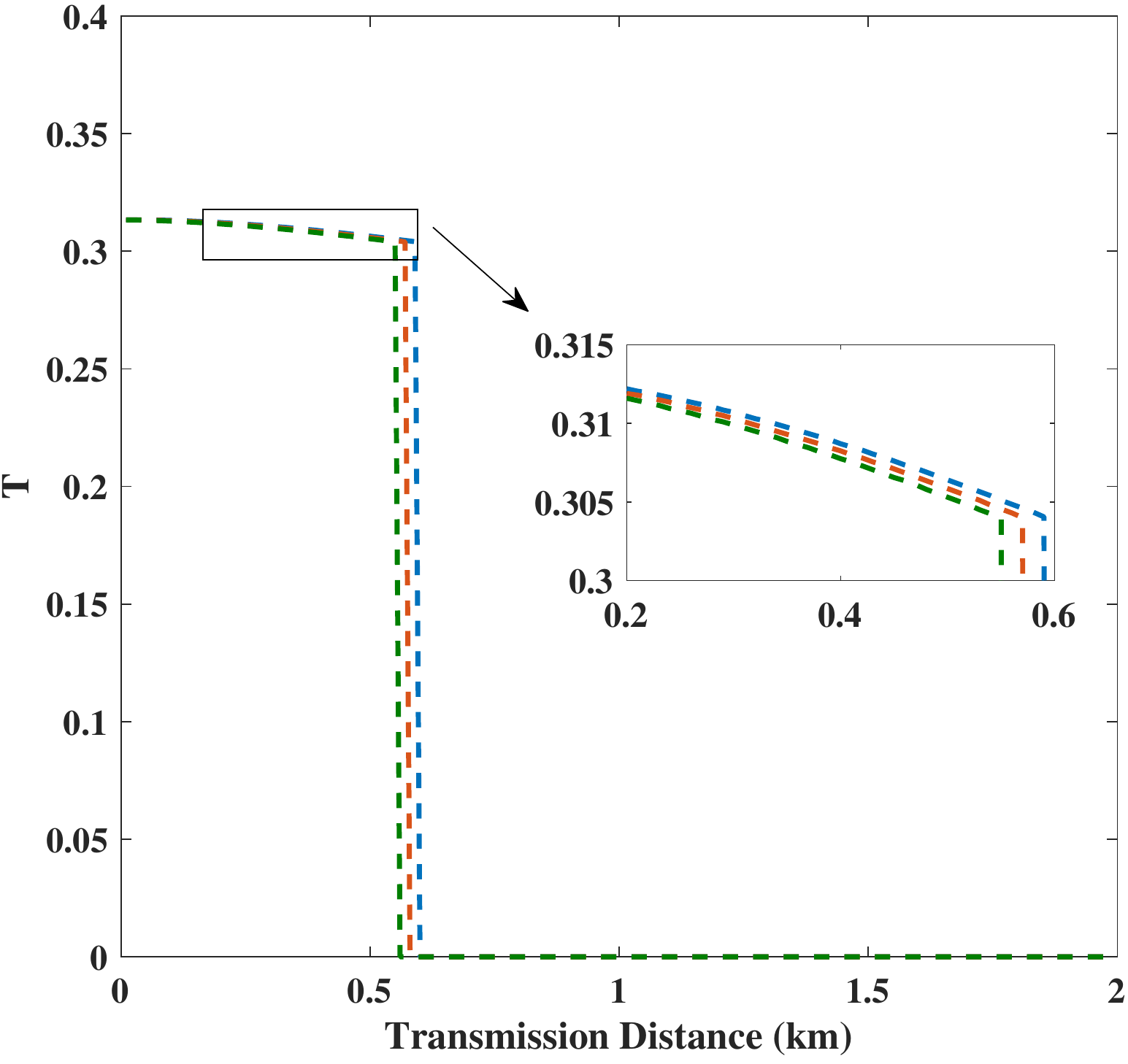}
\label{Fig3d}
}
\caption{{}(Color online) The secret key rate versus the transmission
distance for (a) the extreme asymmetric case and (b) the symmetric case with
several different excess noises $\protect \xi _{j}=\protect \xi \in \left \{
0.002,0.0025,0.003\right \} $ $(j=A,B)$ when optimizing over the
transmittance $T$ shown in (c) and (d), respectively. Note that ZPC Protocol
and Original Protocol represent the ZPC-involved DM MDI-CVQKD and the
original DM MDI-CVQKD. Other parameters are fixed as follows: reconciliation
efficiency is $\protect \beta =0.95,$ optimal variances $V$ for the original
protocol is $1.4$ in the extreme asymmetric case ($1.5$ in the symmetric
case)$,$ optimal variances $V$ for the ZPC protocol is $2.5$ in the extreme
asymmetric case ($2.6$ in the symmetric case).}
\label{Fig3}
\end{figure*}

\begin{figure*}[t]
\centering
\subfigure[]{
\centering
\includegraphics[width=0.47\linewidth]{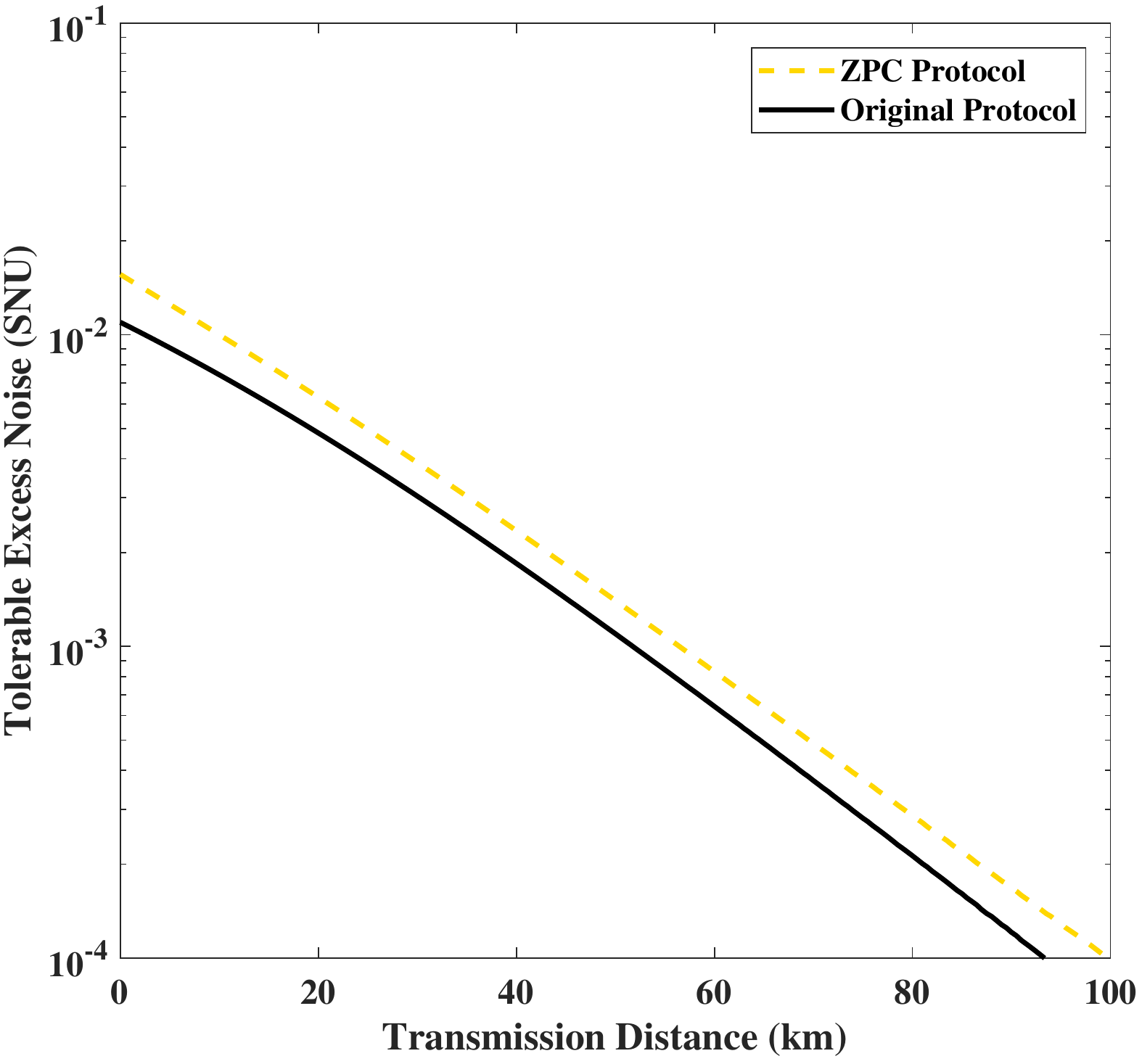}
\label{Fig4a}
}
\subfigure[]{
\includegraphics[width=0.465\linewidth]{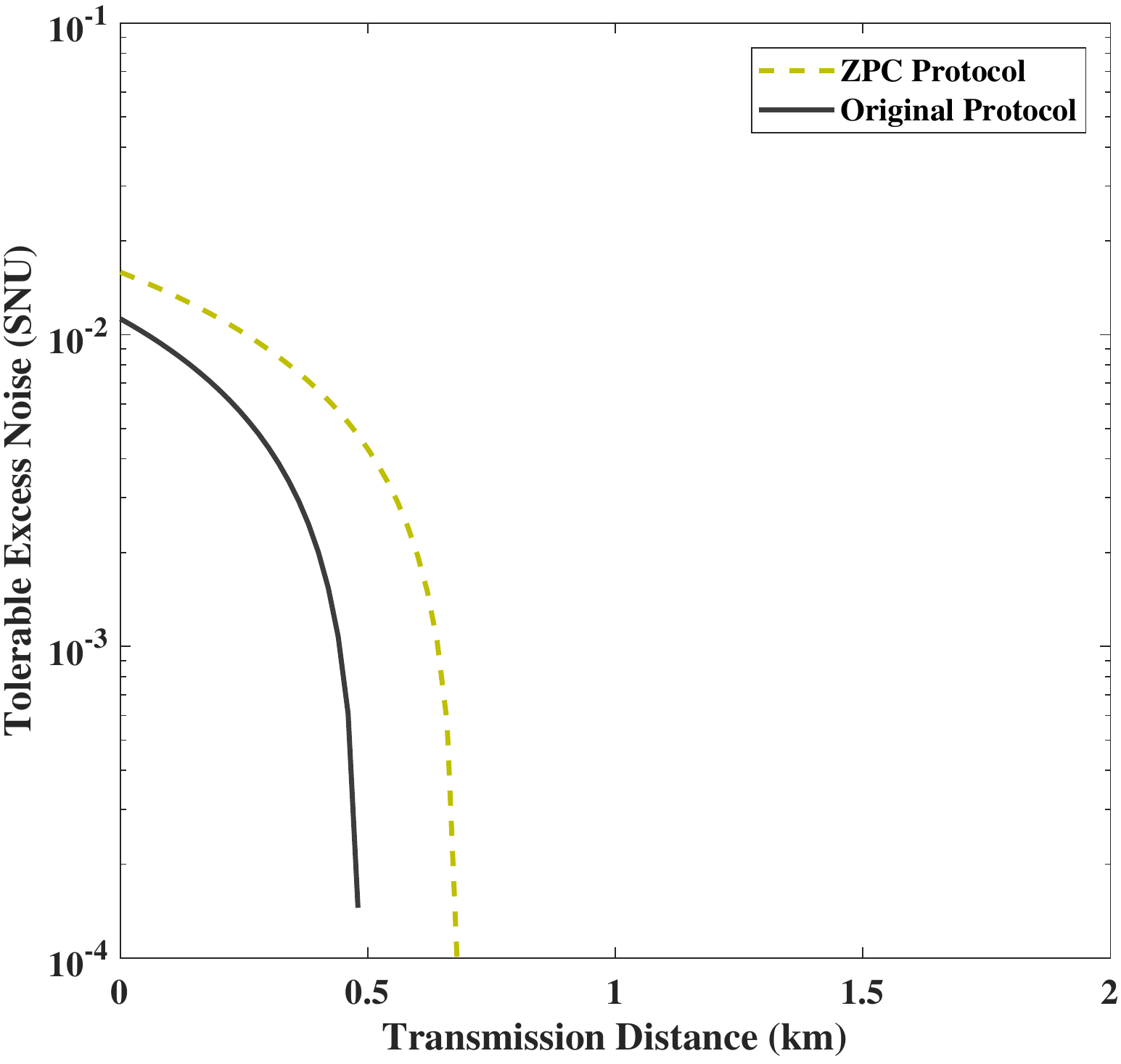}
\label{Fig4b}
}
\caption{{}(Color online) The tolerate excess noise versus the transmission
distance for (a) the extreme asymmetric case and (b) the symmetric case when
optimized over the transmittance $T$. Note that ZPC Protocol and Original
Protocol represent the ZPC-involved DM MDI-CVQKD and the original DM
MDI-CVQKD. Other parameters are fixed as follows: reconciliation efficiency
is $\protect \beta =0.95,$ optimal variances $V$ for the original protocol is
$1.4$ in the extreme asymmetric case ($1.5$ in the symmetric case)$,$
optimal variances $V$ for the ZPC protocol is $2.5$ in the extreme
asymmetric case ($2.6$ in the symmetric case). }
\label{Fig4}
\end{figure*}

\begin{figure*}[t]
\centering
\subfigure[]{
\centering
\includegraphics[width=0.47\linewidth]{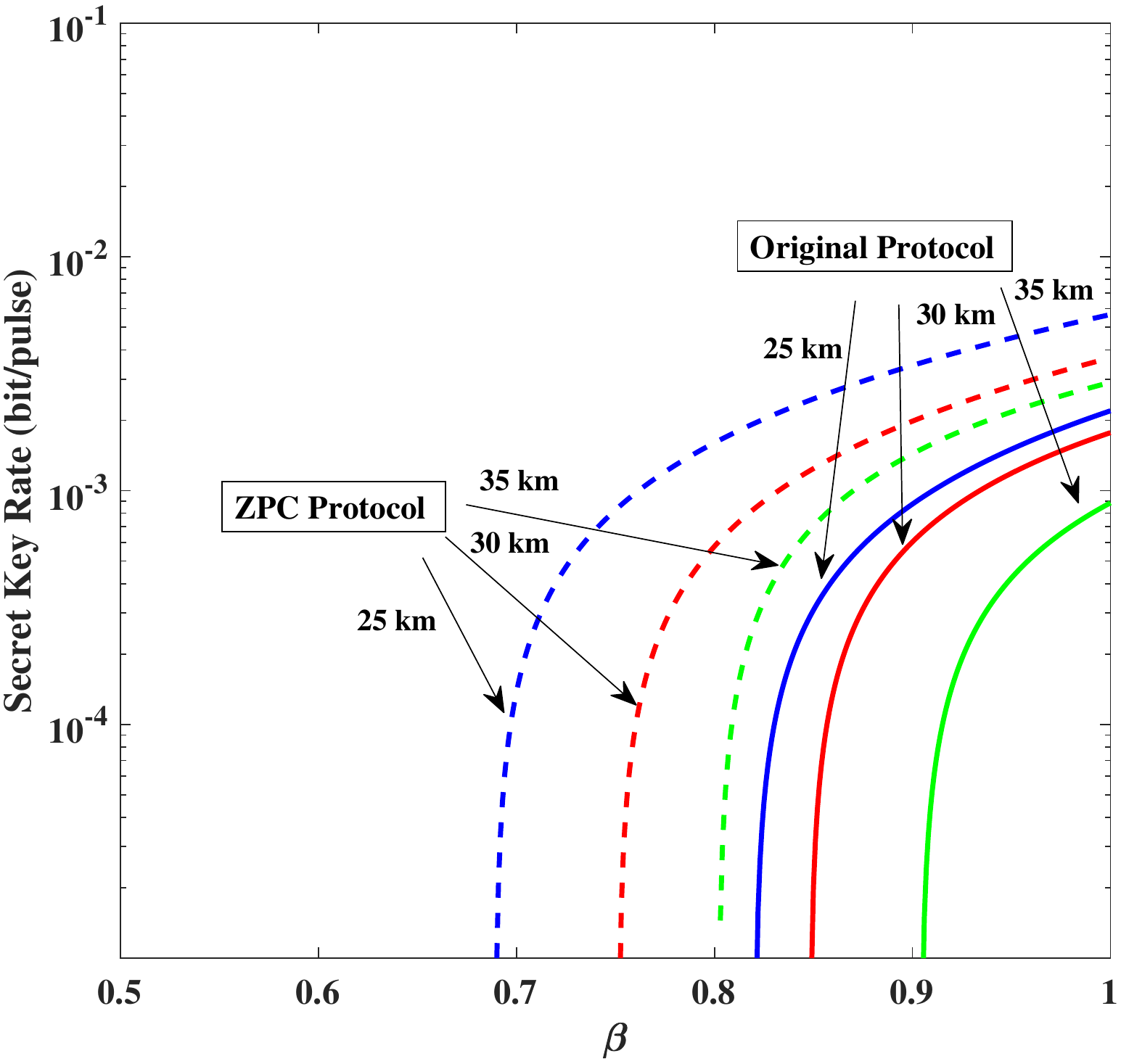}
\label{Fig5a}
}
\subfigure[]{
\includegraphics[width=0.47\linewidth]{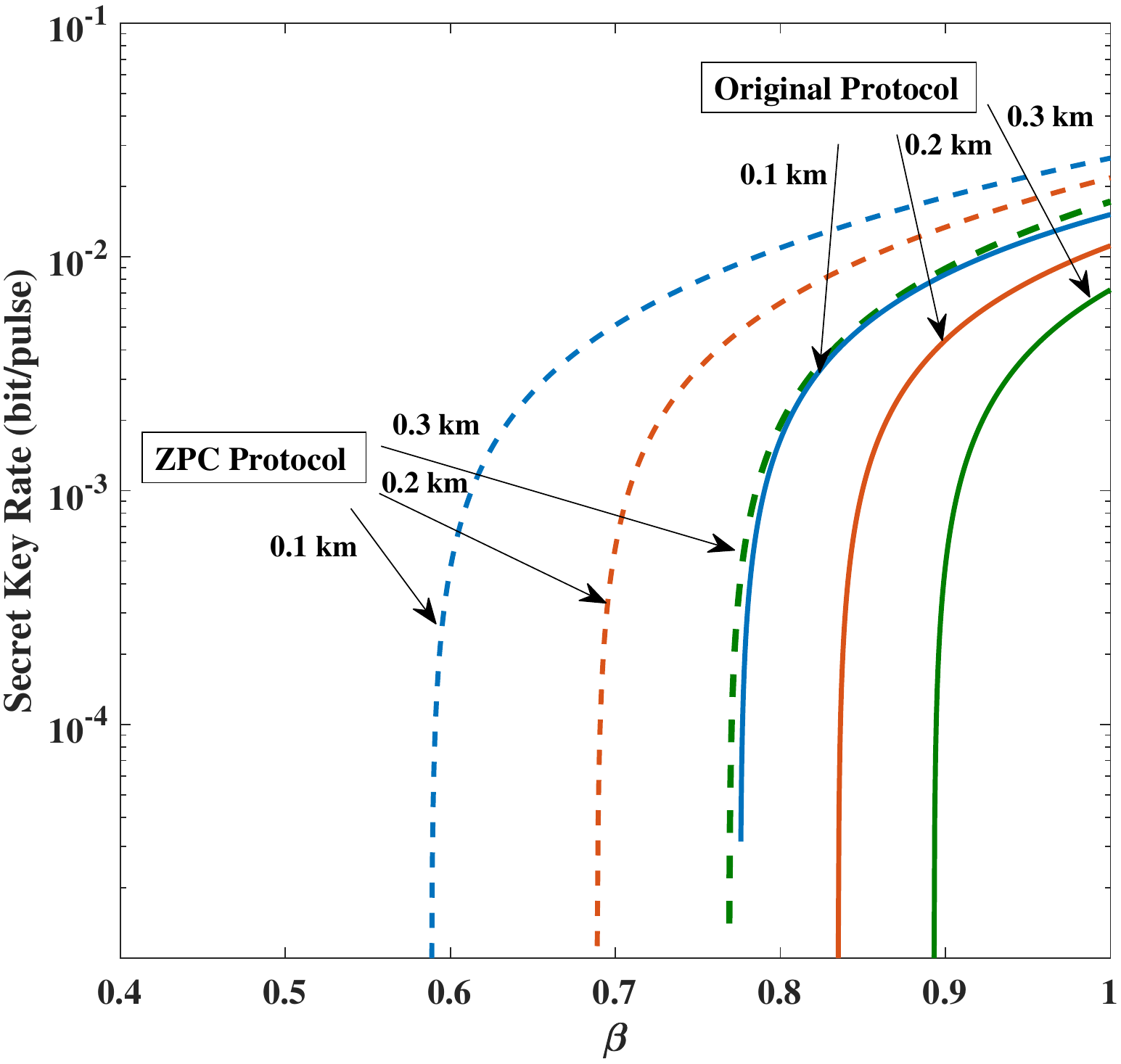}
\label{Fig5b}
}
\caption{{}(Color online) The secret key rate versus the reconciliation
efficiency $\protect \beta $ for (a) the extreme asymmetric case and (b) the
symmetric case with with several different transmission distances e.g., $25$%
km, $30$km, $35$km and $0.1$km $0.2$km, $0.3$km when optimized over the
transmittance $T$, respectively. Note that ZPC Protocol and Original
Protocol represent the ZPC-involved DM MDI-CVQKD and the original DM
MDI-CVQKD. Other parameters are fixed as follows: excess nosie is $\protect%
\xi _{j}=\protect \xi =0.002$ $(j=A,B),$ optimal variances $V$ for the
original protocol is $1.4$ in the extreme asymmetric case ($1.5$ in the
symmetric case)$,$ optimal variances $V$ for the ZPC protocol is $2.5$ in
the extreme asymmetric case ($2.6$ in the symmetric case). }
\label{Fig5}
\end{figure*}

\section{Numerical simulations and discussions}

In the traditional MDI-CVQKD system, most of investigations show that the
extreme asymmetric case has the best performance when comparing with the
symmetric case ($L_{AC}=L_{BC}$) \cite{20,21,22,23,24,25,26,27,33,34,35}.
That is, if Charlie is close to Bob, then the extreme asymmetric case $%
\left( L_{BC}=0\right) $ is more suitable for the point-to-point
communication in contrast to the symmetric case, enabling us to achieve the
maximal transmission distance. Different from the former, the latter has an
unique short-distance network application, e.g., the quantum repeater where
the relay (Charlie) has to be placed in the middle of the legitimate
communication parties. To extract the maximal secret key rates as many as
possible, in the following, we demonstrate the optimal area and value of the
variance $V$ that is an important factor of affecting the performance of
MDI-CVQKD systems with discrete modulation, and then proceed with the
security analysis, including the extreme asymmetric and symmetric cases.

\subsection{Parameter optimization}

Now, let us examine the optimal area and value of $V$ in the ZPC-involved DM
MDI-CVQKD system, which determines the transmitting power of the quantum
signal, thereby impacting the performance of the CVQKD systems. We note that
the actual modulated variance $\widetilde{V}_{M}$ of the proposed system is
expressed as $\widetilde{V}_{M}=\widetilde{V}-1=T\left( V-1\right) $\
according to the relation of $\widetilde{\alpha }=\sqrt{T}\alpha $, which
means that the value of $V$ can be expanded appropriately by using the ZPC
operation on DM MDI-CVQKD. To understand this point, for the given $\beta
=0.95$, Figs. 2 (a) and 2(b) show the secret key rates as a function of the
variance $V$ with the extreme asymmetric and symmetric cases, when
optimizing over the transmittance $T$ shown in Figs. 2(c) and 2(d),
respectively. As for the original DM MDI-CVQKD protocols (solid line),
including the extreme asymmetric and symmetric cases, the optimal areas of $%
V $ are gradually compressed with the decrease of the transmission distance,
whereas for the ZPC-involved DM MDI-CVQKD protocols (dashed line), the
secret key rate decreases much more slowly with the increase of $V$. In
other words, the ZPC-involved protocols have much larger optimal areas of $V$
than the original protocols, which implies that employing the ZPC operation
on the DM MDI-CVQKD systems would bring about more flexible and stable
performances. We also notice that, when $V$ exceeds a certain threshold, the
secret key rate of the ZPC-involved DM MDI-CVQKD system can be higher than
that of the original one, which reveals that the usage of the ZPC operation
can offer a possible way to increase the secret key rate well. To be
specific, for the original protocols, the optimal value of $V$ under the
extreme asymmetric case (the symmetric case) is about $1.4$ ($1.5$) with
several different transmission distances, e.g., $25$km, $30$km and $35$km ($%
0.1$km, $0.2$km and $0.3$km), as described in Ref. \cite{32}. While for the
proposed protocols, we find that the correspondingly optimal $V$ under the
extreme asymmetric case (the symmetric case) is about $2.5$ ($2.6$). As
shown in Figs. 2(c) and 2(d), the corresponding transmittances are,
respectively, given by $T\in \{0.275,0.266,0.258\}$ and $T\in
\{0.312,0.311,0.310\}$ so that the modulated variance $\widetilde{V}_{M}$ of
the proposed protocols can also satisfy the constraint $\widetilde{V}%
_{M}<0.5 $. Thus, in the following, we shall show the best performance of
the DM MDI-CVQKD system when taking into account these optimal variances.

\subsection{Security analysis}

The security of the traditional QKD systems can be reflected from three
aspects, i.e., the secret key rate, the tolerable excess noise and the
transmission distance. In what follows, we focus on the security analysis of
the ZPC-involved DM MDI-CVQKD system, including the extreme asymmetric and
symmetric cases, compared with the original DM MDI-CVQKD.

In Figs. 3(a) and 3(b), we illustrate the secret key rate of both protocols
versus the transmission distance with the special two cases, i.e., the
extreme asymmetric case and the symmetric case, optimized over the
transmittance $T$ depicted in Figs. 3(c) and 3(d), respectively. The solid
lines denote that the original protocols, which can be surpassed by the
protocols using the ZPC operation in terms of the secret key rate and the
transmission distance. Compared with the original protocol, the transmission
distance of the proposed protocol is closer to the PLOB bound \cite{52} that
represents the ultimate limit of repeaterless communications. One reason is
that, by regulating the transmittance $T$, the optimal variance $V$ for the
latter can be obtained in the long-distance communication, and can be bigger
than the former case ($V=1.4$ under the extreme asymmetric case and $V=1.5$
under the symmetric case). The other is that the ZPC operation can be indeed
viewed as a noiseless linear attenuation, which has been used for improving
the transmission distance \cite{48}. In addition, when comparing with the
extreme asymmetric case, the performance of both protocols in the symmetric
case is very poor under the same parameters, because the gap of the excess
noise between the extreme asymmetric and symmetric cases becomes large with
the increase of the transmission distance \cite{32}. We find that the
performances of the above DM MDI-CVQKD systems, even with the ZPC operation,
are affected by the increase of $\xi =0.002,0.0025$ and $0.003$, especially
in terms of the maximal transmission distance. It reveals that the
above-mentioned protocols are sensitive to the excess noise. To further
understand this point, Figs. 4(a) and 4(b) show the tolerable excess noise
as a function of the transmission distance with the extreme asymmetric and
symmetric cases, when taking into account each possible transmittance $T$.
We find that, with the help of the ZPC operation, the tolerable excess noise
of the proposed protocol can be further improved, which implies that under
the same acceptable excess noise, the ZPC-involved DM MDI-CVQKD protocol has
obvious advantages over the original protocol in terms of the maximal
transmission distance. In particular, when $\xi \approx 0.001$, the proposed
protocol is able to support a robust DM MDI-CVQKD system over long distance
about $56$km. However, there is a serious problem that, as for the extreme
asymmetric and symmetric cases, once the excess noise of both protocols is
greater than a certain threshold (about $0.015$ for the proposed protocol
and $0.011$ for the original protocol), we can not achieve the positive
secret key rate. To solve this problem, fortunately, a new two-way CVQKD to
improve robustness against excess noise has been proposed \cite{49}. We also
notice that, with the increase of transmission distance, the excess noise of
both protocols in the symmetric case falls faster than that in the extreme
asymmetric one, thereby making these protocols perform worse than the
extreme asymmetric case with respect to long-distance communication.

The reconciliation efficiency $\beta $, on the other hand,\ is an important
indicator of extracting secret key information. As shown in Figs 5(a) and
5(b), we give the secret key rate versus the reconciliation efficiency $%
\beta $ with the extreme asymmetric and symmetric cases, respectively. We
find that, in the extreme asymmetric case, the available range of $\beta $
for both the proposed protocol (dashed lines) and the original protocol
(solid lines) narrows as the transmission distance increases. Moreover, for
the given transmission distance, the performance of the ZPC-involved DM
MDI-CVQKD system is always better than that of the original one with respect
to the secret key rate and the tolerant of $\beta $. This phenomenon may be
caused by two aspects that the optimal $V$ can be legitimately expanded by
using the ZPC operation on the DM MDI-CVQKD system (shown in Figs. 2(a) and
2(b)) and the system with discrete modulation works well at the extremely
low signal-to-noise ratio with a high-efficiency error correction code in
the reconciliation process. These results are also true for the symmetric
case, as illustrated in Fig. 5(b). In this sense, it reveals that compared
with the original protocol, the ZPC-involved DM MDI-CVQKD protocol in the
symmetric case is more beneficial for short-distance communication,
especially under the low reconciliation efficiency.

\section{Conclusion}

We have suggested the performance improvement of the MDI-CVQKD with discrete
modulation by performing the ZPC operation. We focus on the conventional
four-state scheme as the representation of the DM-based CVQKD system. Due to
the benefits of a high-efficiency error correction code in the
reconciliation process, this four-state-based scheme may offers a possible
way to extend the maximal transmission distance. In the context of an
asymptotic-regime security analysis including the extreme asymmetric and
symmetric cases, our results under the same accessible parameters show that
the secret key rate of the ZPC-involved DM MDI-CVQKD protocol can be
increased, compared with the original protocol. In addition, the performance
of the ZPC-involved protocol is always superior to that of the original one
in terms of the maximal transmission distance. Furthermore, we find that our
protocol enables the DM MDI-CVQKD system to tolerate more lower
reconciliation efficiency. However, the tolerable excess noise of the both
protocols, especially for the symmetric case, decreases with the increase of
the transmission distance, which restrains the effects of extending the
security transmission distance. Fortunately, the two-way CVQKD protocol to
tolerate more excess noise than the one-way protocol was first proposed by
S. Pirandola \cite{49}, which paves the way for our future investigations in
the development of lengthening the security transmission distance.

\begin{acknowledgments}
This work was supported by the National Natural Science Foundation of China
(Grant Nos. 61572529, 61821407, 11964013, 11664017), the Outstanding Young
Talent Program of Jiangxi Province (Grant No. 20171BCB23034), the Training
Program for Academic and Technical Leaders of Major Disciplines in Jiangxi
Province, the Postgraduate Scientific Research Innovation Project of Hunan
Province (Grant No. CX20190126) and the Postgraduate Independent Exploration
and Innovation Project of Central South University (Grant No. 2019zzts070).
\end{acknowledgments}


\bigskip

\bigskip

\bigskip

\bigskip

\begin{thebibliography}{99}
\bibitem{1} H. K. Lo, M. Curty and K. Tamaki, Secure quantum key
distribution, Nat. Photonics \textbf{8}, 595--604 (2014).

\bibitem{2} B. Korzh, C. C. W. Lim, R. Houlmann, N. Gisin, M. J. Li, D.
Nolan, B. Sanguinetti, R. Thew and H. Zbinden, Provably secure and practical
quantum key distribution over 307 km of optical fibre, Nat. Photonics
\textbf{9}, 163--168 (2015).

\bibitem{3} N. Gisin, Gregoire Ribordy, Wolfgang Tittel, and Hugo Zbinden,
Quantum cryptography, Rev. Mod. Phys. \textbf{74}, 145 (2002).

\bibitem{4} S. L. Braunstein and P. van Loock, Quantum information with
continuous variables, Rev. Mod. Phys. \textbf{77}, 513 (2005).

\bibitem{5} V. Scarani, H. Bechmann-Pasquinucci, N. J. Cerf, M. Du\v{s}ek,
N. Lutkenhaus, and M. Peev, The security of practical quantum key
distribution, Rev. Mod. Phys. \textbf{81}, 1301 (2009).

\bibitem{6} A. K. Ekert, Quantum cryptography based on Bell's theorem, Phys.
Rev. Lett. \textbf{67}, 661 (1991).

\bibitem{7} C. Weedbrook, S. Pirandola, R. Garcia-Patron, N. J. Cerf, T. C.
Ralph, J. H. Shapiro, and S. Lloyd, Gaussian quantum information, Rev. Mod.
Phys. \textbf{84}, 621 (2012).

\bibitem{8} P. Jouguet, and S. Kunz-Jacques, and A. Leverrier, and Grangier,
P. and E. Diamanti, Experimental demonstration of long-distance
continuous-variable quantum key distribution, Nat. Photonics. \textbf{7},
378--381 (2013).

\bibitem{9} F. Grosshans and P. Grangier, Continuous variable quantum
cryptography using coherent states, Phys. Rev. Lett. \textbf{88}, 057902
(2002).

\bibitem{10} D. Huang, P. Huang, D. K. Lin, C. Wang, and G. H. Zeng,
High-speed continuous-variable quantum key distribution without sending a
local oscillator, Opt. Lett. \textbf{40}, 3695-3698 (2015).

\bibitem{11} Y. J. Shen, X. Peng, J. Yang, and H. Guo, Continuous-variable
quantum key distribution with Gaussian source noise, Phys. Rev. A \textbf{83}%
, 052304 (2011).

\bibitem{57} B. Qi, P. Lougovski, R. Pooser, W. Grice, and M. Bobrek,
Generating the Local Oscillator \textquotedblleft Locally\textquotedblright \
in Continuous-Variable Quantum Key Distribution Based on Coherent Detection,
Phys. Rev. X \textbf{5}, 041009 (2015).

\bibitem{12} R. G. Patron and N. J. Cerf, Unconditional optimality of
Gaussian attacks against continuous-variable quantum key distribution,\
Phys. Rev. Lett. \textbf{97}, 190503 (2006).

\bibitem{13} F. Furrer, T. Franz, M. Berta, A. Leverrier, V. B. Scholz, M.
Tomamichel, and R. F. Werner, Continuous variable quantum key distribution:
Finite-key analysis of composable security against coherent attacks, Phys.
Rev. Lett. \textbf{109}, 100502 (2012).

\bibitem{14} Y, C. Zhang, Z. Y. Li, Z. Y. Chen, C. Weedbrook, Y. J. Zhao, X.
Y. Wang, Y. D. Huang, C. C. Xu, X. X. Zhang, Z. Y. Wang, M. Li, X. Y. Zhang,
Z. Y. Zheng, B. J. Chu, X. Y. Gao, N. Meng, W. W. Cai, Z. Wang, G. Wang, S.
Yu, and H. Guo, Continuous-variable QKD over 50 km commercial fiber, Quantum
Sci. Technol. \textbf{4}, 035006 (2019).

\bibitem{15} X. C. Ma, S. H. Sun, M. S. Jiang, and L. M. Liang, Local
oscillator fluctuation opens a loophole for Eve in practical
continuous-variable quantum-key-distribution systems, Phys. Rev. A \textbf{88%
}, 022339 (2013).

\bibitem{16} X. C. Ma, S. H. Sun, M. S. Jiang, and L. M. Liang, Wavelength
attack on practical continuous-variable quantum-key-distribution system with
a heterodyne protocol, Phys. Rev. A \textbf{87}, 052309 (2013).

\bibitem{17} J. Z. Huang, C. Weedbrook, Z. Q. Yin, S. Wang, H. W. Li, W.
Chen, G. C. Guo, and Z. F. Han, Quantum hacking of a continuous-variable
quantum-key-distribution system using a wavelength attack, Phys. Rev. A
\textbf{87}, 062329 (2013).

\bibitem{18} H. K. Lo, M. Curty, and B. Qi, Measurement-device-independent
quantum key distribution, Phys. Rev. Lett. \textbf{108}, 130503 (2012).

\bibitem{19} S. Pirandola, C. Ottaviani, G. Spedalieri C. Weedbrook, S. L.
Braunstein, S. Lloyd, T. Gehring, C. S. Jacobsen, and U. L. Andersen,
High-rate measurement-device-independent quantum cryptography, Nat. Photon.
\textbf{9}, 397 (2015).

\bibitem{20} X. Y. Zhang, Y. C. Zhang, Y. J. Zhao, X. Y. Wang, S. Yu, and H.
Guo, Finite-size analysis of continuous-variable
measurement-device-independent quantum key distribution, Phys. Rev. A
\textbf{96}, 042334 (2017).

\bibitem{21} Z. Y. Li, Y. C. Zhang, F. H. Xu, X. Peng, and H. Guo,
Continuous-variable measurement-device-independent quantum key distribution,
Phys. Rev. A \textbf{89}, 052301 (2014).

\bibitem{22} X. C. Ma, S. H. Sun, M. S. Jiang, M. Gui, and L. M. Liang,
Gaussian-modulated coherent-state measurement-device-independent quantum key
distribution, Phys. Rev. A 89, 042335 (2014).

\bibitem{23} C. Lupo, C. Ottaviani, P. Papanastasiou, and S. Pirandola,
Continuous-variable measurement-device-independent quantum key distribution:
Composable security against coherent attacks, Phys. Rev. A 97, 052327 (2018).

\bibitem{24} Y. C. Zhang, Z. Y. Li, S. Yu, W. Y. Gu, X. Peng, and H. Guo,
Continuous-variable measurement-device-independent quantum key distribution
using squeezed states, Phys. Rev. A \textbf{90}, 052325 (2014).

\bibitem{25} X. Y. Zhang, Y. C. Zhang, Y. J. Zhao, X. Y. Wang, S. Yu, and H.
Guo, Finite-size analysis of continuous-variable
measurement-device-independent quantum key distribution, Phys. Rev. A
\textbf{96}, 042334 (2017).

\bibitem{26} P. Papanastasiou, C. Ottaviani, and S. Pirandola, Finite-size
analysis of measurement-device-independent quantum cryptography with
continuous variables, Phys. Rev. A \textbf{96}, 042332 (2017).

\bibitem{27} Z. Y. Chen, Y. C. Zhang, G. Wang, Z. Y. Li, and H. Guo,
Composable security analysis of continuous-variable
measurement-device-independent quantum key distribution with squeezed states
for coherent attacks, Phys. Rev. A \textbf{98}, 012314 (2018).

\bibitem{56} Y. D. Wu, J. Zhou, X. B. Gong, Y. Guo, Z. M. Zhang, and G. Q.
He, Continuous-variable measurement-device-independent multipartite quantum
communication, Phys. Rev. A \textbf{93}, 022325 (2016).

\bibitem{33} H. X. Ma, P. Huang, D. Y. Bai, S. Y. Wang, W. S. Bao, and G. H.
Zeng, Continuous-variable measurement-device-independent quantum key
distribution with photon subtraction, Phys. Rev. A \textbf{97}, 042329
(2018).

\bibitem{34} Y. J. Zhao, Y. C. Zhang, B. J. Xu, S. Yu, and H. Guo,
Continuous-variable measurement-device-independent quantum key distribution
with virtual photon subtraction, Phys. Rev. A \textbf{97}, 042328 (2018).

\bibitem{35} P. Wang, X. Y. Wang, and Y. M. Li, Continuous-variable
measurement-device-independent quantum key distribution using modulated
squeezed states and optical amplifiers, Phys. Rev. A \textbf{99}, 042309
(2019).

\bibitem{36} Y. Guo, W. Ye, H. Zhong, and Q. Liao, Continuous-variable
quantum key distribution with non-Gaussian quantum catalysis, Phys. Rev. A
\textbf{99}, 032327 (2019).

\bibitem{37} W. Ye, H. Zhong, Q. Liao, D. Huang, L. Y. Hu, and Y. Guo,
Improvement of self-referenced continuous-variable quantum key distribution
with quantum photon catalysis, Opt. Express \textbf{27}, 17186-17198 (2019).

\bibitem{38} L. Y. Hu, J. N. Wu, Z. Y. Liao, and M. S. Zubairy, Multiphoton
catalysis with coherent state input: Nonclassicality and decoherence, J.
Phys. B: At. Mol. Phys. \textbf{49}, 175504 (2016).

\bibitem{39} A. I. Lvovsky and J. Mlynek, Quantum-optical catalysis:
generating nonclassical states of light by means of linear optics, Phys.
Rev. Lett. \textbf{88}, 250401 (2002).

\bibitem{40} W. D. Zhou, W. Ye, C. J. Liu, L. Y. Hu, and S. Q. Liu,
Entanglement improvement of entangled coherent state via multiphoton
catalysis, Laser Phys. Lett. \textbf{15}, 065203 (2018).

\bibitem{41} L. Y. Hu, Z. Y. Liao, and M. S. Zubairy, Continuous-variable
entanglement via multiphoton catalysis, Phys. Rev. A \textbf{95}, 012310
(2017).

\bibitem{42} W. Ye, H. Zhong, X. D. Wu, L. Y. Hu, and Y. Guo,
Continuous-variable measurement-device-independent quantum key distribution
via quantum catalysis, arXiv:1907.03383v2 (2019).

\bibitem{43} T. J. Richardson, M. A. Shokrollahi, and R. Urbanke, Design of
capacity-approaching irregular low-density parity-check codes, IEEE Trans.
Inf. Theory \textbf{47}, 619 (2001).

\bibitem{44} A. Leverrier, R. Alleaume, J. Boutros, G. Zemor, and P.
Grangier, Multidimensional reconciliation for a continuous-variable quantum
key distribution, Phys. Rev. A \textbf{77}, 042325 (2008).

\bibitem{28} A. Leverrier and P. Grangier, Unconditional security proof of
long-distance continuous-variable quantum key distribution with discrete
modulation, Phys. Rev. Lett. \textbf{102}, 180504 (2009).

\bibitem{29} A. Leverrier and P. Grangier, Continuous-variable
quantum-key-distribution protocols with a non-Gaussian modulation,\ Phys.
Rev. A \textbf{83}, 042312 (2011).

\bibitem{30} P. Huang, J. Fang, and G. H. Zeng, State-discrimination attack
on discretely modulated continuous-variable quantum key distribution, Phys.
Rev. A \textbf{89}, 042330 (2014).

\bibitem{31} Y. Shen, H. X. Zou, L. Tian, P. X. Chen, and J. M. Yuan,
Experimental study on discretely modulated continuous-variable quantum key
distribution, Phys. Rev. A \textbf{82}, 022317 (2010).

\bibitem{32} H. X. Ma, P. Huang, D. Y. Bai, T. Wang, S. Y. Wang, W. S. Bao,
and G. H. Zeng, Long-distance continuous-variable
measurement-device-independent quantum key distribution with discrete
modulation, Phys. Rev. A \textbf{99}, 022322 (2019).

\bibitem{51} S. Ghorai, P. Grangier, E. Diamanti, and A. Leverrier,
Asymptotic security of continuous-variable quantum key distribution with a
discrete modulation, Phys. Rev. X \textbf{9}, 021059 (2019).

\bibitem{54} W. Ye, Y. Guo, Y. Xia, H. Zhong, H. Zhang, J. Z. Ding, and L.
Y. Hu, Discrete modulation continuous-variable quantum key distribution
based on quantum catalysis, Acta Phys. Sin. \textbf{69}, 060301 (2020).

\bibitem{55} J. Yang, B. J. Xu, X. Peng, and H. Guo, Four-state
continuous-variable quantum key distribution with long secure distance,
Phys. Rev. A \textbf{85}, 052302 (2012).

\bibitem{45} M. Navascues, F. Grosshans, and A. Acin, Optimality of Gaussian
attacks in continuous-variable quantum cryptography, Phys. Rev. Lett.
\textbf{97}, 190502 (2006).

\bibitem{46} S. Pirandola, Entanglement reactivation in separable
environments,\ New J. Phys. \textbf{15}, 113046 (2013).

\bibitem{50} H. Zhang and G. Q. He, Improving the performance of the
four-state continuous-variable quantum key distribution by using optical
amplifiers, Phys. Rev. A \textbf{86}, 022338 (2012).

\bibitem{53} Y. C. Zhang, Z. Y. Li, C. Weedbrook, S. Yu, W. Y. Gu, M. Z.
Sun, X. Peng, and H. Guo, Improvement of two-way continuous-variable quantum
key distribution using optical amplifiers, J. Phys. B: At. Mol. Opt. Phys.
\textbf{47}, 035501 (2014).

\bibitem{47} M. M. Wolf, G. Giedke, and J. I. Cirac, Extremality of Gaussian
quantum states, Phys. Rev. Lett. \textbf{96}, 080502 (2006).

\bibitem{52} S. Pirandola, R. Laurenza, C. Ottaviani, and L. Banchi,
Fundamental limits of repeaterless quantum communications, Nat. Commun.
\textbf{8}, 15043 (2017).

\bibitem{48} J. Fiurasek and N. J. Cerf, Gaussian postselection and virtual
noiseless amplification in continuous-variable quantum key distribution.
Phys. Rev. A \textbf{86}(6), 060302(R) (2012).

\bibitem{49} S. Pirandola, S. Mancini, S. Lloyd, and S. L. Braunstein,
Continuous-variable quantum cryptography using two-way quantum
communication, Nat. Phys. \textbf{4}, 726--730 (2008).
\end{thebibliography}
\end{document}